# Nonlinear PDE Constrained Optimal Dispatch of Gas and Power: A Global Linearization Approach

Yuan Li, *Student Member, IEEE,* Shuai Lu, Wei Gu, *Senior Member, IEEE,*
Yijun Xu, *Senior Member, IEEE*, Ruizhi Yu, Suhan Zhang, Zhikai Huang

*Abstract*—The coordinated dispatch of power and gas in the electricity-gas integrated energy system (EG-IES) is fundamental for ensuring operational security. However, the gas dynamics in the natural gas system (NGS) are governed by the nonlinear partial differential equations (PDE), making the dispatch problem of the EG-IES a complicated optimization model constrained by nonlinear PDE. To address it, we propose a globally linearized gas network model based on the Koopman operator theory, avoiding the commonly used local linearization and spatial discretization. Particularly, we propose a data-driven Koopman operator approximation approach for the globally linearized gas network model based on the extended dynamic mode decomposition, in which a physics-informed stability constraint is derived and embedded to improve the generalization ability and accuracy of the model. Based on this, we develop an optimal dispatch model for the EG-IES that first considers the nonlinear gas dynamics in the NGS. The case study verifies the effectiveness of this work. Simulation results reveal that the commonly used locally linearized gas network model fails to accurately capture the dynamic characteristics of NGS, bringing potential security threats to the system.

## I. Introduction

### A. Motivations

WITH the continuous growth of global energy demand and the urgent demand for environmental sustainability, integrating different energy sectors is believed to be a promising solution for improving energy efficiency and promoting renewable energy source consumption, receiving more and more attention from academia and industry [1]. Particularly, with the increase of gas-fired generators and power-to-gas (P2G) units, the coupling between the natural gas system (NGS) and electrical power system (EPS) is becoming tighter, posing significant challenges for the system operation [2, 3].

The coordinated dispatch of the NGS and EPS in the electricity-gas integrated energy system (EG-IES) plays a crucial role in the operational security and economy. First, by coordinating electricity and natural gas, the complementarity between them in production, transmission, and consumption can be utilized, thereby improving overall safety and energy efficiency [4, 5]. Second, since the NGS and EPS are coupled by some units, such as the gas-fired generators and P2G units, the operational conditions of the two systems need to be well coordinated to ensure security [6]. However, the gas dynamics in the pipeline are governed by the nonlinear partial differential equations (PDE), including the momentum and mass conservation equations. Therefore, the dispatch model of the EG-IES is typically a complicated nonlinear PDE constrained optimization problem, which is very challenging to solve. Lots of research has tried to address this problem by using a steady gas model that ignores the gas dynamics in the pipeline or using a linearized dynamic model [6, 7], which however, cannot accurately describe the operational states of the NGS, bringing significant threat to the security (e.g., the system states exceed the limit). Considering this, this paper aims to address the nonlinear PDE constrained dispatch problem of the EG-IES based on a globally linearized gas network model.

### B. Related Work

In existing research, two categories of gas network models are used in the dispatch of EG-IES, including the steady-state and dynamic models. The steady-state gas model uses the algebraic equations (AE), i.e., the Weymouth equation, to describe the gas state under equilibrium. This model ignores the gas dynamics in the pipelines by assuming that the gas flow can achieve equilibrium instantaneously, which is often used in the planning problem. Saldarriaga et al. [8] present a holistic approach for the co-planning of EPS and NGS based on the steady-state gas network model. Zhang et al. [9] propose a multi-stage stochastic planning scheme for the EG-IES, in which all stages utilize the steady-state gas network model. Besides, the steady state gas network model is sometimes used in the day-ahead dispatch problem, for example, [10] and [11]. Note that the comparative analysis in various studies [12, 13] points out that considering the dynamic characteristics of NGS is essential for the dispatch of EG-IES despite its higher computational complexity.

The dynamic gas network model uses nonlinear PDE to describe the dynamic behavior of gas flow in the pipeline, which presents two challenges for embedding it into the dispatch models, including linearization and discretization. The average velocity-based linearization method is widely adopted in existing research. For example, Fang *et al.* [14] established the optimal energy flow model of EG-IES based on the average velocity-based dynamic gas model and compared it with the steady-state gas model. Notably, the accuracy of this model is influenced by the choice of the average flow velocity [15], leading to a lack of general applicability. To address this issue, some studies have discussed how to modify the average flow velocity based on the operational conditions to improve the model accuracy [16]. However, in the dispatch model, the

This work was supported in part by the National Natural Science Foundation of China under Grant 52325703, and in part by the National Key R&D Plan Project of China under Grant 2022YFB2404000 *(Corresponding author: Shuai Lu.)*

operational conditions of the gas network are part of decision variables, indicating that it is challenging to select appropriate average flow velocities a priori that can ensure the accuracy of the model.

Furthermore, it is necessary to discretize the linearized PDE into the AE so that the dispatch model can resort to the classical numerical algorithms [16]. The primary approach for this problem is the finite difference method, such as the Euler difference scheme [17] and the Wendroff difference scheme [18]. These discretization methods can produce potential numerical issues, such as numerical dispersion and dissipation. Also, the difference method introduces large-scale additional variables and constraints that represent the states of different nodes in time and space, greatly increasing the computational burden. An effective method to avoid discretization is the functional space transformation, typically including Laplace transformation [19, 20], Fourier transformation [21], and Bernstein space transformation [22]. Usually, the functional space transformation method introduces extra errors and computational burdens to handle the initial conditions of the NGS, sometimes outweighing the benefits and requiring careful consideration of their applicability. More importantly, this method can only deal with the linearized PDE model.

In summary, the research gaps in the dispatch of the EG-IES include: (1) Almost all the existing research adopts a linearized dynamic gas model in the dispatch of EG-IES, failing to capture the nonlinear gas dynamics adequately; (2) There is still no effective method reported in existing research that can effectively address the nonlinear PDE constrained dispatch problem of the EG-IES.

C. *Contributions and Paper Organization*

To bridge the above research gaps, this paper first proposes a global linearization approach for the nonlinear PDE model of the NGS based on the Koopman operator theory. Then, an optimal dispatch model that considers the nonlinear gas dynamics of the NGS is developed for the EG-IES. To the authors' knowledge, this is the first work to effectively address the nonlinear PDE constrained dispatch problem of the EG-IES.

The main contributions are summarized as follows.
1) We first propose a globally linearized gas network model based on the Koopman operator theory. By this, we convert the original nonlinear PDE model of the gas network into linear AE, which avoids the spatial difference and can be easily integrated into the dispatch model.
2) We propose a data-driven Koopman operator approximation approach for the globally linearized gas network model based on the extended dynamic mode decomposition (EDMD), in which a physics-informed stability constraint is derived and embedded to improve the generalization ability and accuracy of the model.
3) We propose an optimal dispatch model for the EG-IES that considers the nonlinear gas dynamics in the NGS. By comparative analysis, we reveal that the commonly used locally linearized PDE model fails to accurately capture the dynamic characteristics of NGS, bringing potential security threats to the system.

The remainder of this paper is as follows. Section II introduces the optimal dispatch model of the EG-IES. Section III derives the globally linearized model of the NGS based on the Koopman operator theory. Section IV presents the Koopman operator approximation method. Case studies are given in Section V, and Section VI concludes this work.

## II. OPTIMAL DISPATCH MODEL

In this section, we first introduce the optimal dispatch model of the EG-IES. Then, the challenges caused by the nonlinear PDE constraints of the gas pipeline are analyzed.

A. *Model Formulation*

The optimal dispatch of EG-IES aims to determine the device output and power flow of the network to minimize the prescribed objective function. The concise form of the optimal dispatch model of EG-IES is as

$$\begin{aligned} \min_{x_e, x_g} \quad & f(x_e, x_g) \\ s.t. \quad & h_e(x_e) \leq 0 \\ & h_g(x_g) \leq 0 \\ & h_c(x_e, x_g) = 0 \end{aligned} \quad . \tag{0}$$

In model (0), $x_e$ denotes the decision variables of the EPS, including the bus voltage magnitude, the voltage angle, the active/reactive power flow at branches, and the power of generators; $x_g$ denotes the decision variables of the NGS, including the pressure and mass flow rate (MFR) of pipelines, and the MFR at the gas source; $f(\cdot)$ represents the objective function. The EPS constraints $h_e(\cdot) \leq 0$ include the power flow equations, the transmission capacity limits, the phase angle limits, the supply and demand power balance at each period, and the power limits of units. Here, we use the commonly used DC power flow to model the electrical network. The NGS constraints $h_g(\cdot) \leq 0$ include the gas dynamic equations of the pipeline, the node balance constraints, the MFR limitations, and pressure limitations [14]. Besides, the coupling constraints $h_c(\cdot) \leq 0$ include the energy conversion relationship of the coupling devices, such as the gas-fired generators and P2G units. A detailed optimal dispatch model of EG-IES refers to [12, 14].

The gas dynamic equations of the pipeline are critical constraints of the NGS. Based on the mass conservation [23] and the momentum conservation [24], the gas dynamics in the pipeline are described by

$$\frac{\partial \rho}{\partial t} + \frac{\partial (\rho v)}{\partial x} = 0, \tag{1a}$$

$$\frac{\partial (\rho v)}{\partial t} + \frac{\partial (\rho v^2)}{\partial x} + \frac{\partial p}{\partial x} + g(\rho - \rho_a)\sin\alpha + \frac{\lambda \rho v^2}{2d} = 0, \tag{1b}$$

wherein $\rho$ is the gas density, $v$ is the gas flow velocity, $p$ is the pressure, $g$ is the gravitational acceleration, $\rho_a$ is the reference density, $\alpha$ is the pipeline inclination angle, $\lambda$ is the friction factor, and $d$ is the diameter of the pipeline.

The mass conservation equation (1a) ensures that mass is neither created nor destroyed within the pipeline. The momentum equation (1b) captures the various forces acting on a gas

parcel within the pipeline. Besides, the state equation links the pressure and density as

$$p = c^2 \rho, \quad (1c)$$

wherein $c$ is the speed of sound.

The following assumptions and simplifications are made on the gas dynamic equations. First, since the fluid velocity $v$ is much smaller than the speed of sound $c$, the convective term $\partial(\rho v^2)/\partial x$ always remains at a small value under standard operating conditions and thus can be neglected [16]. Second, it is assumed that the altitude along the pipeline is constant (i.e., $\alpha = 0$), so the term $g(\rho - \rho_\alpha)\sin\alpha$ in (1b) can be omitted. Besides, the fluid transport in the pipeline is assumed to be isothermal, meaning the gas temperature is constant, and thus $c^2$ in (1c) is also constant.

Furthermore, we introduce the MFR, which can be defined by $M = \rho v A$. Then, by substituting $\rho = p/c^2$ and $\rho v = M/A$ into (1a) and (1b), the gas dynamic equations (1a)~(1c) can be transformed into the following ones

$$\begin{cases} \dfrac{\partial p}{\partial t} + \dfrac{c^2}{A}\dfrac{\partial M}{\partial x} = 0 \\ \dfrac{\partial p}{\partial x} + \dfrac{1}{A}\dfrac{\partial M}{\partial t} + \dfrac{\lambda c^2 M^2}{2dA^2 p} = 0 \end{cases}, \quad (2)$$

wherein $A$ is the cross-sectional area of the pipeline, and $M$ is the MFR.

B. *Challenges Caused by Nonlinear PDE Constraints*

The nonlinear PDE (2) in the constraints $h_g(x_g) \leq 0$ makes the optimal dispatch model (0) a nonlinear PDE constrained optimization problem, posing a significant challenge to the solution. The existing studies usually use the average velocity-based linearization method to handle the nonlinear term in the momentum equation. In this method, by assuming $v \approx \overline{v}$, we have

$$\frac{\lambda v^2 \rho}{2d} \approx \frac{\lambda \overline{v} v \rho}{2d} = \frac{\lambda \overline{v} M}{2dA}, \quad (3a)$$

wherein $\overline{v}$ is the average flow velocity.

Then, the momentum equation in (2) can be linearized as

$$\frac{\partial p}{\partial x} + \frac{1}{A}\frac{\partial M}{\partial t} + \frac{\lambda \overline{v} M}{2dA} = 0. \quad (3b)$$

Apparently, the average velocity-based method is a typical local linearization technique, the accuracy of which largely depends on the preselected average flow velocity $\overline{v}$. Here, we analyze the simulation results for a pipeline under various average velocities. Assuming the parameters $L$, $d$, $\lambda$, and $c$ are respectively 30km, 0.5m, 0.0108, and 340m/s. The prescribed inlet pressure is set to 5.78MPa, with the outlet flow declining from 15kg/s to 10kg/s at $t = 1$h. The stable backward Euler method with a time step of 0.25h is employed for the simulation. Fig. 1 (a) and (b) show the inlet flow and outlet pressure under different average velocities, using the simulation results of the nonlinear PDE (2) as the reference. Obviously, the values of $\overline{v}$ have a considerable influence on the internal state of the pipeline, and the final stable value of pressure varies notably under different $\overline{v}$. However, in the optimal dispatch prob-lem, it is unrealistic to select a proper average flow velocity $\overline{v}$ a prior since the MFR and pressure of the gas in the NGS are decision variables. Hence, the errors caused by the average velocity-based linearized method will bring unavoidable risks to the operational safety and economy of the EG-IES.

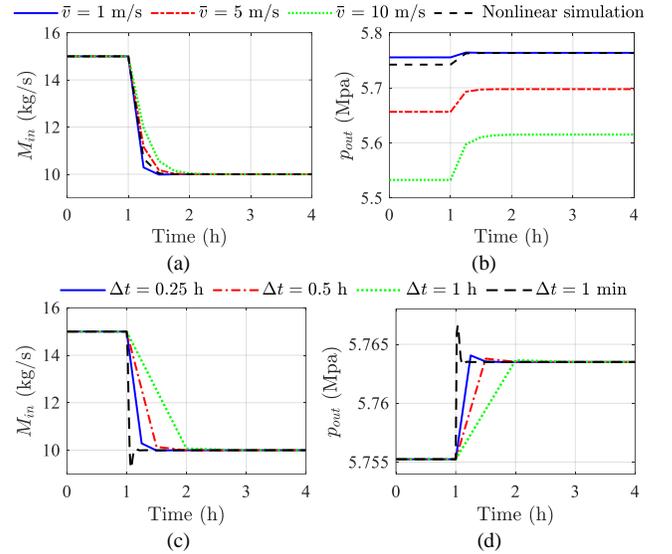

Fig. 1. Simulation results under different average velocities and time steps: (a) Inlet MFR at different average velocities; (b) Outlet pressure at different average velocities; (c) Pressure dynamic with different time steps; (d) MFR dynamic with different time steps.

To investigate the impact of different time steps on the model accuracy, we employ the backward Euler method under various time step schemes at the same $\overline{v} = 1$m/s, and the results are given in Fig. 1 (c) and (d). The results indicate that the gas dynamics all end after about 1 hour, and the final stable values of the pressure and MFR remain consistent under different time steps. Also, because of the existence of numerical oscillation, mindlessly reducing the time step may not improve the accuracy of the results. Thus, selecting a proper time step is important for balancing the accuracy and computational complexity.

***Remark 1***. In summary, using the average velocity-based linearized model of the gas network in the dispatch of EG-IES brings the following issues:
1) Average flow velocity selection: Linearization methods usually require a pre-set average flow velocity $\overline{v}$, but this is difficult to achieve in gas pipelines where mass flow and pressure constantly change.
2) Model accuracy: The average velocity-based method might have large deviations compared to the more accurate nonlinear PDE model, especially in pressure, posing potential security risks.
3) Numerical stability: Usually, we need to discrete the PDE gas model into AE, in which choosing proper spatial and temporal steps is critical because an improper step can lead to large numerical errors or instability.

III. GLOBALLY LINEARIZED MODEL OF GAS NETWORK

In this section, we propose the data-driven globally linearized model of the gas network based on the Koopman operator theory. First, we introduce the ODE representation of the gas pipeline. Second, we present the lifting-based global linearization method. Third, we derive the input-output linear model of the gas pipeline using the delay-embedded technique to avoid the spatial difference.

### A. ODE Representation of Gas Pipeline

Since the Koopman operator theory applies to the ODE system, we use the semi-discrete method to convert the nonlinear PDE of the gas pipeline into the ODE. First, we discrete the pipeline into $K$ elements. Then, for the $k$th element, we can discretize the spatial differential terms $\partial M/\partial x$ and $\partial p/\partial x$ of and obtain the corresponding ODE approximation, as

$$\begin{cases} \dot{p}_{out}^k = g_1\left(M_{out}^k, M_{in}^k\right) \\ \dot{M}_{out}^k = g_2\left(M_{out}^k, p_{out}^k, M_{in}^k, p_{in}^k\right) \end{cases}, \quad (4)$$

wherein $M_{in}^k$ and $M_{out}^k$ are the inlet and outlet MFR, $p_{in}^k$ and $p_{out}^k$ are the inlet and outlet pressure, and $g_1(\cdot)$ and $g_2(\cdot)$ are the equations depending on the specific spatial difference scheme. For example, using the explicit Euler scheme $\partial M^k/\partial x \approx (M_{out}^k - M_{in}^k)/\Delta x$, $\partial p^k/\partial x \approx (p_{out}^k - p_{in}^k)/\Delta x$ and under the approximation $M^2/p \approx \frac{1}{2}(M_{out}^k + M_{in}^k)^2/(p_{out}^k + p_{in}^k)$, we have

$$\begin{cases} \dot{p}_{out}^k = -\dfrac{c^2}{A}\dfrac{M_{out}^k - M_{in}^k}{\Delta x} \\ \dot{M}_{out}^k = -A\dfrac{p_{out}^k - p_{in}^k}{\Delta x} - \dfrac{\lambda c^2 \left(M_{out}^k + M_{in}^k\right)^2}{4dA\left(p_{out}^k + p_{in}^k\right)} \end{cases}.$$

Note that the proposed method in the following does not need to specify a discretization method. By defining $x^k = [M_{out}^k \ p_{out}^k]^\top \in \mathbb{R}^2$, $u^k = [M_{in}^k \ p_{in}^k] \in \mathbb{R}^2$, the equation (4) can be recast as

$$\dot{x}^k = g(x^k, u^k). \quad (5a)$$

The above equations indicate that the gas dynamics in the pipeline is a nonlinear dynamic system with control. For convenience, we use the discrete-time form of (5a), as

$$x^{k,t} = g\left(x^{k,t-1}, u^{k,t}, u^{k,t-1}\right). \quad (5b)$$

### B. Lifting-Based Global Linearization Method

The principle of Koopman operator theory in a single pipeline is shown in Fig. 2. Based on the Koopman operator theory, we can lift the system in the nonlinear space $\mathcal{F}_o$ into an infinite-dimensional linear space $\mathcal{F}$ by the Koopman observables $\psi$, as the process indicated by the red arrow in Fig. 2. Then, the nonlinear system (5b) in $\mathcal{F}_o$ can be transformed into a linear system in the infinite-dimensional space $\mathcal{F}$. Mathematically, we have

$$\psi\left(x^{k,t}\right) = \psi \circ g\left(x^{k,t-1}\right) = \mathcal{K}_x^k \psi\left(x^{k,t-1}\right), \quad (6a)$$

wherein $\circ$ represents the composition operation, $\mathcal{K}_x^k$ is the infinite Koopman operator; $\psi$ is also infinite-dimensional theoretically and can be represented as $\psi(x^{k,t}) = \left[M_{out}^{k,t} \ p_{out}^{k,t} \ \psi_M(M_{out}^{k,t})^\top \ \psi_p(p_{out}^{k,t})^\top\right]^\top$.

To integrate the control variables $u^{k,t}$, we extend (6a) into

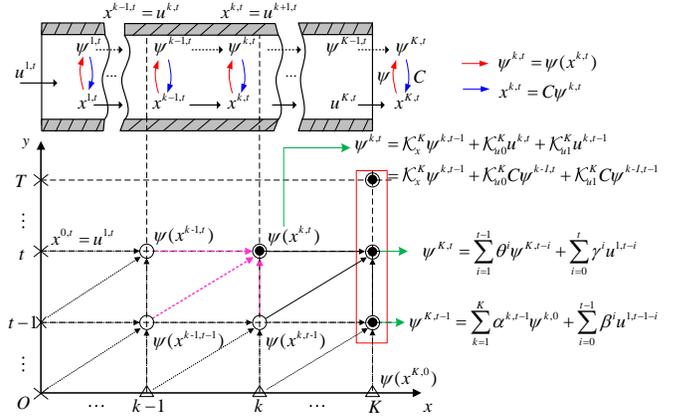

Fig. 2. Schematic of the pipeline recursive model.

the following one

$$\psi\left(x^{k,t}\right) = \mathcal{K}_x^k \psi\left(x^{k,t-1}\right) + \mathcal{K}_{u0}^k u^{k,t} + \mathcal{K}_{u1}^k u^{k,t-1}, \quad (6b)$$

wherein $\mathcal{K}_{u0}^k$, $\mathcal{K}_{u1}^k$ are the infinite linear operators denoting the impact of $u^{k,t}$ and $u^{k,t-1}$.

### C. Input-Output Model of the Pipeline

We choose the state variables $x^{k,t}$ as the first observable, and thus, we can easily get the original state variables in $\mathcal{F}_o$, as the process presented by the blue arrow in Fig. 2. The mathematical expression can be written as

$$x^{k,t} = \begin{bmatrix} M_{out}^{k,t} & p_{out}^{k,t} \end{bmatrix}^\top = C\psi\left(x^{k,t}\right), \quad (7a)$$

wherein $C = [I_{2\times 2} \ 0_{2\times \infty}] \in \mathbb{R}^{2\times \infty}$.

As shown in Fig. 2, the output variables of the $k$th element in the pipeline, $u^{k,t}$, is the input of the $(k+1)^{th}$ element, i.e.,

$$u^{k,t} = x^{k-1,t} = C\psi\left(x^{k-1,t}\right) \quad 2 \le k \le K. \quad (7b)$$

The schematic diagram of the pipeline is also shown in Fig. 2. The left boundary condition signifies the input at the initial segment of the pipeline (marked by ×), while the lower boundary condition denotes the initial state variables at the ends of the pipeline (marked by △). The target is to obtain the outflow pressure and MFR profiles in the high-dimensional space ($\psi(x^{K,t}), 1 < t < T$). It is obvious in Fig. 2 that $\psi(x^{k,t})$ can be represented by a linear combination of states from the surrounding three nodes, as denoted by ○ in Fig. 2.

Recursing forward from the $K$th segment, we have

$$\begin{aligned} \psi\left(x^{K,t}\right) &= \mathcal{K}_x^K \psi\left(x^{K,t-1}\right) + \mathcal{K}_{u0}^K u^{K,t} + \mathcal{K}_{u1}^K u^{K,t-1} \\ &= \left(\mathcal{K}_x^K\right)^t \psi\left(x^{K,0}\right) \\ &\quad + \sum_{i=1}^{t} \left(\mathcal{K}_x^K\right)^{i-1} \left(\mathcal{K}_{u0}^K u^{K,t+1-i} + \mathcal{K}_{u1}^K u^{K,t-i}\right) \\ &= \left(\mathcal{K}_x^K\right)^t \psi\left(x^{K,0}\right) + \mathcal{K}_{u0}^K u^{K,t} \\ &\quad + \sum_{i=1}^{t-1} \left(\left(\mathcal{K}_x^K\right)^i \mathcal{K}_{u0}^K + \left(\mathcal{K}_x^K\right)^{i-1} \mathcal{K}_{u1}^K\right) u^{K,t-i} \\ &\quad + \left(\mathcal{K}_x^K\right)^{t-1} \mathcal{K}_{u1}^K u^{K,0} \end{aligned}. \quad (8a)$$

Based on (7b), we have

$$\psi\left(x^{K,t}\right) = \left(\mathcal{K}_x^K\right)^t \psi\left(x^{K,0}\right) + \mathcal{K}_{u0}^K C\psi\left(x^{K-1,t}\right)$$
$$+ \sum_{i=1}^{t-1}\left(\left(\mathcal{K}_x^K\right)^i \mathcal{K}_{u0}^K + \left(\mathcal{K}_x^K\right)^{i-1}\mathcal{K}_{u1}^K\right)C\psi\left(x^{K-1,t-i}\right)$$
$$+ \left(\mathcal{K}_x^K\right)^{t-1}\mathcal{K}_{u1}^K C\psi\left(x^{K-1,0}\right)$$
$$= \sum_{k=1}^{K}\alpha^{k,t}\psi\left(x^{k,0}\right) + \sum_{i=0}^{t}\beta^i u^{1,t-i} \qquad (8b)$$

wherein the coefficients $\alpha^{k,t}$ and $\beta^i$ are constants composed of $\mathcal{K}_x^k$, $\mathcal{K}_{u0}^k$, and $\mathcal{K}_{u1}^k$.

Thus, the left and lower boundary conditions determine the outlet state. Similarly, the outlet state prior to time $t$ can be described as

$$\psi\left(x^{K,t-j}\right) = \sum_{k=1}^{K}\alpha^{k,t-j}\psi\left(x^{k,0}\right) + \sum_{i=0}^{t-j}\beta^i u^{1,t-j-i} . \qquad (8c)$$
$$j = 1,\cdots,t-1$$

It is evident that the components of $\psi(x^{K,t-j}), j = 1,\ldots,t-1$ are encompassed within $\psi(x^{K,t})$.

Furthermore, when $t \geq K+1$, we can introduce the parameters $\theta^i, i = 1,\ldots,t-1$ satisfying the equations in (8d).

$$\begin{bmatrix} \alpha^{1,1} & \alpha^{1,2} & \cdots & \alpha^{1,t-1} \\ \alpha^{2,1} & \alpha^{2,2} & \cdots & \alpha^{2,t-1} \\ \vdots & \vdots & \ddots & \vdots \\ \alpha^{K,1} & \alpha^{K,2} & \cdots & \alpha^{K,t-1} \end{bmatrix} \begin{bmatrix} \theta^1 \\ \theta^2 \\ \vdots \\ \theta^{t-1} \end{bmatrix} = \begin{bmatrix} \alpha^{1,t} \\ \alpha^{2,t} \\ \vdots \\ \alpha^{K,t} \end{bmatrix} \qquad (8d)$$

Then, the initial constant term in (8b) can be eliminated by the previous states $\psi(x^{K,n}), n = 1, 2, \ldots, t-1$, as

$$\psi\left(x^{K,t}\right) = \sum_{i=1}^{t-1}\theta^i \psi\left(x^{K,t-i}\right) + \sum_{i=0}^{t}\gamma^i u^{1,t-i} , \qquad (8e)$$

wherein $\gamma^i, i = 0, \ldots, t$ are the parameters determined by $\alpha^{k,t-j}$ and $\beta^i$.

This indicates that with enough delay terms, the term $\psi(x^{K,t})$ can be represented by the linear combinations of the control variables at the inlet, $u^{1,i}, i = 0, 1, \cdots, t$, and the outlet state variables at previous moments, $\psi(x^{K,i}), i = 1, 2, \cdots, t-1$.

From the equations (8e), it can be inferred that the number of iterations required for forward recursion of $\psi(x^{K,t-i})$ and $u^{1,t-i}$ increases as $i$ increases. The higher the powers in the coefficients, the more rapidly the effect on the state $\psi(x^{K,t})$ decays. This means that the states and control inputs that are farther away from the current moment contribute less to $\psi(x^{K,t})$.

Finally, we can truncate (8e) to get a concise and approximate model for practical applications as

$$\psi\left(x^{K,t}\right) = \sum_{i=1}^{D_x}\theta^i \psi\left(x^{K,t-i}\right) + \sum_{i=0}^{D_u}\gamma^i u^{1,t-i} + \varepsilon_1^t , \qquad (8f)$$

wherein $D_x$ is the number of delay embeddings of the state variables, $D_u$ is the number of delay embeddings of the input variables, and $\varepsilon_1^t$ denotes the error caused by the truncation of the delay order at moment $t$.

### D. Finite-Dimensional Approximation

Since the Koopman operator is infinite-dimensional, we need a finite-dimensional truncation of the system in practice. Define $\psi_{M,trc} = \begin{bmatrix}\psi_{M,1} & \psi_{M,2} & \cdots & \psi_{M,N_m}\end{bmatrix}^\top$, $\psi_{p,trc} = \begin{bmatrix}\psi_{p,1} & \psi_{p,2} & \cdots & \psi_{p,N_p}\end{bmatrix}^\top$, and $N = N_m + N_p + 2$. Thus, $\mathcal{K}_{trc,x}^k = [\mathcal{K}_x^k]_{1:N} \in \mathbb{R}^{N \times N}$, and $\mathcal{K}_{trc,u}^k = [\mathcal{K}_u^k]_{1:N} \in \mathbb{R}^{N \times 2}$. Then, we can get an $N$ dimensional approximation for the system (6b), as

$$\psi_{trc}\left(x^{k,t}\right) = \mathcal{K}_{trc,x}^k \psi_{trc}\left(x^{k,t-1}\right) + \mathcal{K}_{trc,u0}^k u^{k,t} \\ + \mathcal{K}_{trc,u1}^k u^{k,t-1} + \varepsilon_2^t \qquad (9a)$$

wherein $\varepsilon_2^t$ is introduced to denote the error caused by the finite-dimensional approximation of the Koopman observable.

The system (9a) is a finite approximation of system (6b) with uniform formulation. With the same derivation of (6b) $\sim$(8f), we can get the same form as (8f) in the finite-dimensional space, as

$$\psi_{trc}\left(x^{K,t}\right) = \sum_{i=1}^{D_x}\widetilde{\mathcal{K}}_{trc,x}^i \psi_{trc}\left(x^{K,t-i}\right) + \sum_{i=0}^{D_u}\widetilde{\mathcal{K}}_{trc,u}^i u^{1,t-i} + \varepsilon^t , \qquad (9b)$$

wherein $\widetilde{\mathcal{K}}_{trc,x}^i$ is the finite-dimensional Koopman operator, $\widetilde{\mathcal{K}}_{trc,u}^i$ is the linear operator, and $\varepsilon^t = \varepsilon_1^t + \varepsilon_2^t$ is the total error, namely the "global linearization error".

***Remark 2***. Based on the above results, we clarify two issues:
(1) The derivation in this section indicates that the delayed states of the gas flow need to be embedded into the model to eliminate the spatial difference. The essential reason is that the gas dynamics is described by a PDE instead of an ODE.

(2) Once we get the Koopman observables $\psi_{trc}$, the Koopman operator $\widetilde{\mathcal{K}}_{trc,x}^i$, and the linear operator $\widetilde{\mathcal{K}}_{trc,u}^i$ that can make the global linearization error $\varepsilon^t$ small enough, we can get a linear representation of the gas dynamics of the pipeline. Then, we can replace the original nonlinear PDE constraints of the gas pipeline in the model (0) with the linear constraints (9b) (with $\varepsilon^t = 0$). Note that for one pipeline, the model (9b) only includes one constraint for each time period, which not only provides a globally linear description of the nonlinear gas dynamics but also avoids introducing extra spatial variables as traditional finite difference methods.

### IV. KOOPMAN OPERATOR APPROXIMATION

In this section, we first introduce the approximation method of the Koopman operator. Then, we derive a stability constraint to ensure stable consistency between the systems in Koopman space and the original space.

#### A. High-Dimensional Linearization

We omit the superscript $k$ in the following for conciseness. Assuming we can obtain $M$ sets of snapshots for the state $x^1, x^2, \ldots, x^M$ and control variables $u^1, u^2, \ldots, u^M$. It should be noted that only the state variables $x$ require to be lifted, while the input variables $u$ remain unchanged.

Based on (9b), by omitting the total error, the evolution of the state variables can be expressed as

$$\psi_{trc}\left(x^t\right) \approx \sum_{i=0}^{D_x} \tilde{\mathcal{K}}_{trc,x}^i \psi_{trc}\left(x^{t-i}\right) + \sum_{i=0}^{D_u} \tilde{\mathcal{K}}_{trc,u}^i u^{t-i}, \quad (10)$$

wherein $\tilde{\mathcal{K}}_{trc,x}^0 = 0$.

Based on the assumption that the errors follow a Gaussian distribution, the least squares method can be applied to obtain the Koopman operators $\tilde{\mathcal{K}}_{trc,x}^i$ and $\tilde{\mathcal{K}}_{trc,u}^i$, as

$$\min_{\tilde{\mathcal{K}}_{trc,(\cdot)}^i} \left\| \sum_{t=1}^T \left( \psi_{trc}\left(x^t\right) - \left( \sum_{i=0}^{D_x} \tilde{\mathcal{K}}_{trc,x}^i \psi_{trc}\left(x^{t-i}\right) + \sum_{i=0}^{D_u} \tilde{\mathcal{K}}_{trc,u}^i u^{t-i} \right) \right) \right\|_2^2. \quad (11)$$

Once the Koopman operators are obtained, for any given input, the next state can be predicted in the high-dimensional space by (10).

### B. Stability Constraint

The dissipative nature of the NGS determines that it is a stable system. Accordingly, the system in the Koopman space should have the same property [25]. Therefore, in the Koopman operator approximation, we should add the stability constraints to avoid yielding an unstable system and thus improve the generalization capability and accuracy.

In existing research, the Lyapunov stability constraints are commonly used to ensure the asymptotic stability of systems. Considering delay embedding, the system matrix $A_{sys}$ of the pipeline can be defined as

$$A_{sys} = \begin{bmatrix} \tilde{\mathcal{K}}_{trc,x}^1 & \cdots & \tilde{\mathcal{K}}_{trc,x}^{D_x} \\ \mathbf{I}_{N\cdot(D-1)} & \mathbf{0}_{(N\cdot(D-1))\times N} \end{bmatrix}. \quad (12)$$

To ensure the Lyapunov stability, the mode of system matrix $A_{sys}$ should be strictly less than 1. For $\tilde{\mathcal{K}}_{trc,x}^i, \forall i = 1, \ldots, D_x$, the Lyapunov stability constraint is mathematically expressed as: there exists a Lyapunov function $P > 0$ such that $A_{sys}^T P A_{sys} - \tilde{\rho}^2 P < 0$, wherein $\tilde{\rho}$ is the spectral radius, and it should be set to a number less than 1 to ensure stability. By applying the Schur complement, the inequality can be reformulated as

$$\left(A_{sys}^\top P\right) P^{-1} \left(A_{sys}^\top P\right)^\top - \tilde{\rho}^2 P < 0, \quad (13a)$$

which can be further converted into a positive definite condition as

$$\begin{bmatrix} \tilde{\rho} P & A_{sys}^\top P \\ P^\top A_{sys} & \tilde{\rho} P \end{bmatrix} > 0, \quad P > 0. \quad (13b)$$

Since $A_{sys}$ and $P$ are both variables, (13a) includes a bilinear matrix inequality constraint. This makes the optimization problem nonconvex and NP-hard. Typically, the iterative approach is used to solve this problem, which increases computational complexity and cannot ensure convergence [26]. To address this problem, we propose a new stability constraint, i.e., the spectral radius of the system matrix for a stable system should be less than 1. Mathematically, we have $\rho(A_{sys}) < 1$. However, adding this constraint to the model (11) makes this optimization problem hard to solve. Consequently, we tighten the constraint to ensure that the 2-norm of the system matrix is less than 1. Specifically, we use the following constraint to ensure stability as

$$\text{norm}_2(A_{sys}) \leq 1 - \epsilon \quad i = 1, 2, \ldots, D_x, \quad (14)$$

wherein $\epsilon$ is a small value to make the spectral norm of $A_{sys}$ is strictly less than 1 (set to 0.001 in this work).

Compared to Lyapunov stability constraints, this method avoids the nonconvex BMI constraints, thereby significantly reducing computational complexity.

***Remark 3***. Finally, the proposed stability-constrained Koopman operation approximation model includes the objective function (11) and the stability constraint (14), which is a typical semidefinite programming problem that can be solved by off-the-shelf solvers, such as Mosek and SeDuMi. Theoretically, the tightened stability constraints (14) will make the obtained Koopman operator a suboptimal solution, making the resulting globally linearized model less accurate. However, our numerical results indicate that the loss of accuracy is negligible. More importantly, the original model consisting of (11) and (13b) is nonconvex, making it difficult to find a global optimal solution. Hence, the proposed method here is a good choice.

### C. An Illustrative Case

The same single pipeline tested in Section II-B is used to validate the proposed globally linearized model. In this case, we set $\Delta t = 15\text{min}, D_x = 3, D_u = 2$, the base value is set to $P_b = 5\text{e}^6\text{Pa}, M_b = 10\text{kg/s}$. The inlet pressure and MFR of the pipeline are selected as input variables, while the outlet pressure and MFR are selected as state variables. The variation in the inputs and state variables of all snapshots is given in Fig. 3 (a) and (b). A total of 6400 snapshots were taken, with the first 80% serving as training data and the remaining 20% as test data (denoted by the blue dashed line along the vertical x-axis).

We choose three different observables for the two state variables. The first is $\psi_1(x) = x$, and the other two observables are $\psi_2(x) = -x \cdot e^{-x}$, and $\psi_3(x) = e^{-x} \cdot \sin(-x)$. The normalized errors of the simulation result for $P_{out}$ and $M_{out}$ are illustrated in Fig. 3 (c). It can be observed that even when the inputs show significant fluctuations, the normalized error of $P_{out}$ can be controlled within 2e-4p.u., while that of $M_{out}$ can be roughly kept within 5e-3p.u. The MFR in the pipeline is more sensitive and can be easily influenced by the fluctuation of the pressure, resulting in larger errors. In addition, the fluctuation ranges of the test and training data errors are similar, indicating better stability. Overall, such errors are rather small for both simulation and practical engineering applications. Fig. 3 (d) gives the results without the proposed stability constraints. Although the error without stability constraints also appears small on the training data, the peak error becomes significantly larger. Moreover, in the test data, there is an increase in error when stability constraints are removed. This indicates that the stability constraints can effectively balance stability and error.

## V. CASE STUDY

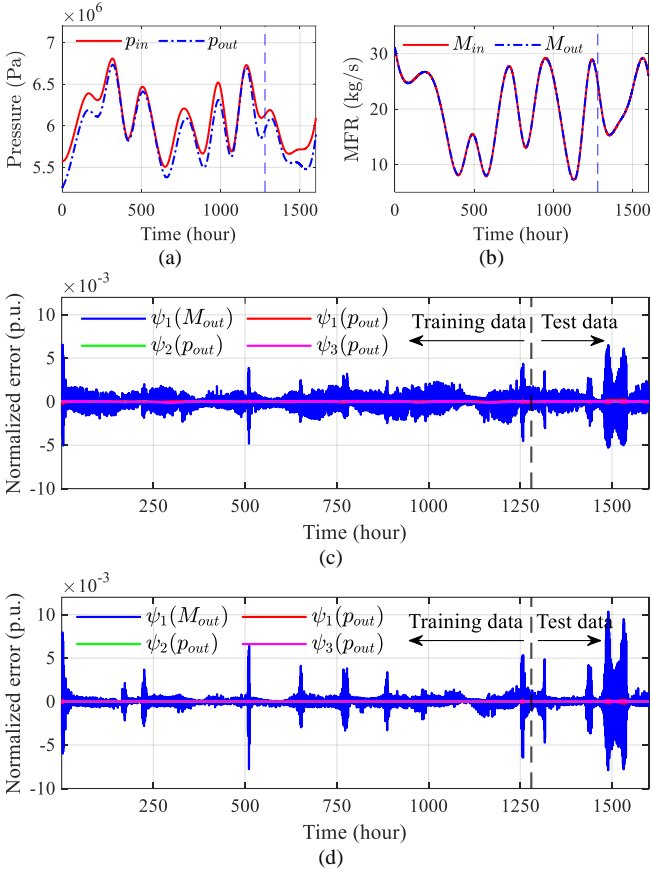

Fig. 3. Relative error of state variables: (a) Pressure; (b) MFR;(c) Error with the stability constraint; (d) Error without the stability constraint.

To demonstrate the effectiveness of the proposed method, we perform the simulations on an EG-IES consisting of an IEEE 30-bus electrical power network and a 20-node gas network modified from the Belgium system [27]. The EPS has 6 generators, including 2 wind turbines, 3 coal-fired generators, and 1 gas-fires generator. The power and gas networks are coupled through a P2G unit and a gas turbine. The system structure is shown in Fig. 4. The detailed parameters are provided in [28]. In the dispatch model, the dispatch interval of the control variables of the gas network, including the pressure at the gas sources and the MFR at the gas loads, is set to 1 hour to keep consistent with the dispatch interval of the EPS. The model resolution of the gas network model is set to 15 minutes to balance the accuracy and computational burden. All the simulations are performed on a laptop with an Intel i7 CPU and 16 GB RAM. The program is coded using Matlab R2022b and YALMIP [29]. We use the Mosek 10.0 to solve the Koopman operator approximation model and the CPLEX 12.10 to solve the optimal dispatch model.

In the following, we first analyze the globally linearized model of the NGS. Second, the error of different models is compared. Finally, we discuss the optimal dispatch results.

### A. Globally Linearized Model of Gas Network

For each pipeline in the NGS, a dataset containing 49800 snapshots is generated by the nonlinear PDE-based simulations, with a resolution of 15 minutes per snapshot. The first 80% of the data is used for training, and the remaining 20% is reserved for testing. The observables chosen for the pipelines are $\psi(x^{k,t}) = \left[ p_{out}^{k,t} \ M_{out}^{k,t} \ -p_{out}^{k,t} \cdot \exp(-p_{out}^{k,t}) \ \exp(-p_{out}^{k,t}) \cdot \sin(-p_{out}^{k,t}) \right]^\top$. The base values are set as $P_b = 5e^6$ Pa and $M_b = 10$ kg/s. Table I presents the error indicators for training and test data under different delay embedding orders. The first four rows show the results when the proposed stability constraints are applied. Clearly, the root mean square error (RMSE) and mean absolute percentage error (MAPE) of the globally linearized model are small. As the embedding delay $D_x$ increases, the RMSE of state variables decreases, improving the accuracy of linear approximation. However, once $D_x$ exceeds 3, the improvements in error indicators become negligible.

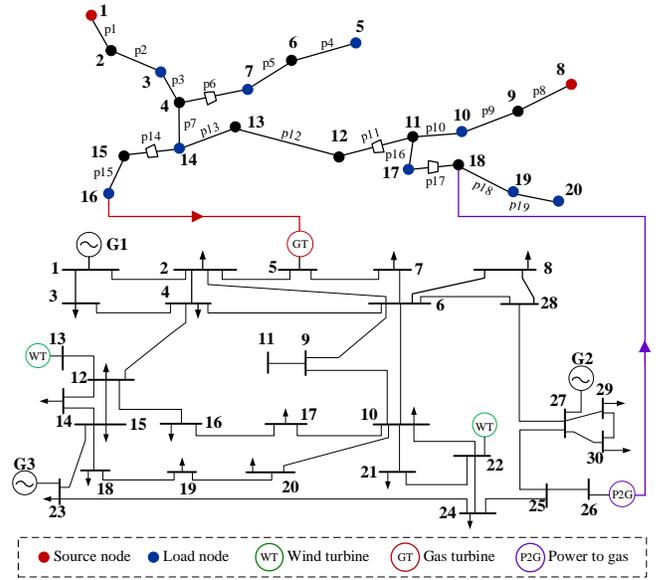

Fig. 4. The structure of the system in Case Study.

TABLE I
ERROR INDICATORS FOR TRAINING DATA

| Model resolution of gas network (min) | | 15 | | | 30 | 60 |
|---|---|---|---|---|---|---|
| $(D_x, D_u)$ | | (2,2) | (3,2) | (4,2) | (3,2) | (3,2) |
| RMSE[1] | $p_{out}$ (Pa) | 306.89 | 274.37 | 276.17 | 380.74 | 908.59 |
| | $M_{out}$ (kg/s) | 0.0065 | 0.0065 | 0.0065 | 0.0271 | 0.0651 |
| MAPE[1] | $p_{out}$ | 0.0017% | 0.0015% | 0.0015% | 0.0026% | 0.0067% |
| | $M_{out}$ | 0.0155% | 0.0154% | 0.0153% | 0.0837% | 0.2027% |
| RMSE[2] | $p_{out}$ (Pa) | 61.93 | 46.89 | 44.70 | 170.26 | 563.81 |
| | $M_{out}$ (kg/s) | 0.0050 | 0.0038 | 0.0037 | 0.0205 | 0.0568 |
| MAPE[2] | $p_{out}$ | 0.0003% | 0.0002% | 0.0002% | 0.0011% | 0.0038% |
| | $M_{out}$ | 0.0120% | 0.0086% | 0.0073% | 0.0548% | 0.1797% |

[1] Error indicators with the stability constraint; [2] Error indicators without the stability constraint.

Table I also displays the error indicators for model resolutions of 30 and 60 minutes, revealing that the error increases as the resolution grows. The last four rows, which exclude stability constraints, show relatively lower errors. This suggests that excluding stability constraints improves the model performance in the training process, as adding constraints imposes additional limitations and increases error. However, our simulations show that applying the Koopman operators trained without stability constraints in the dispatch model leads to numerical issues, making the optimization model unsolvable. This further underlines the significance of considering stability constraints in the training process.

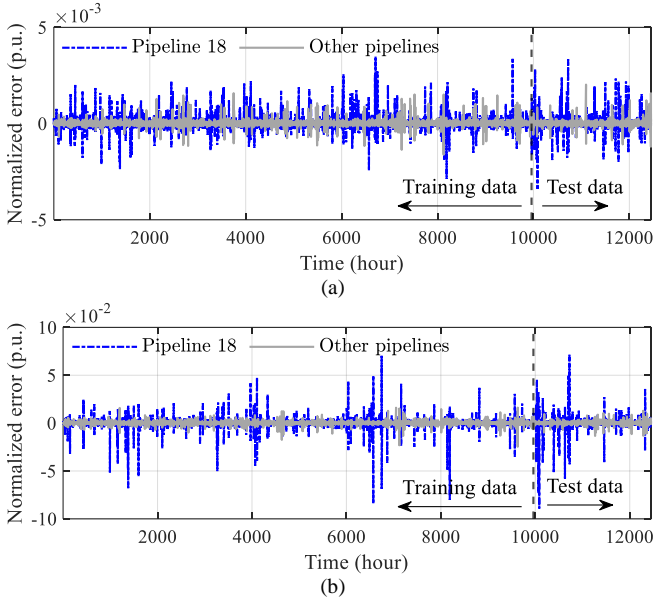

Fig. 5. The normalized error of the training and test data: (a) $P_{out}$; (b) $M_{out}$.

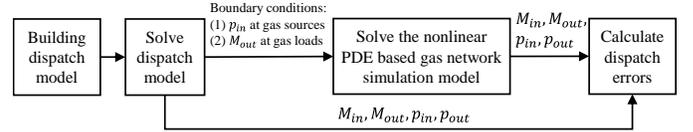

Fig. 6. Flowchart of gas network simulation and error calculation.

TABLE II
NETWORK-LEVEL ERROR OF GLOBALLY LINEARIZED MODEL

| Model resolution of gas network (min) | | 15 | 30 | 60 |
|---|---|---|---|---|
| RMSE | $p_{out}$ (MPa) | 0.00459916 | 0.00458854 | 0.00516590 |
| | $M_{in}$ (kg/s) | 0.118 | 0.206 | 0.319 |
| MAPE | $p_{out}$ | 0.04% | 0.04% | 0.04% |
| | $M_{in}$ | 0.66% | 1.06% | 1.83% |

In the dispatch model, we employ the globally linearized model trained with a 15-minute resolution, where $D_x = 3$, and $D_u = 2$. Fig. 5 shows the normalized errors of $P_{out}$ and $M_{out}$ across all pipelines. The outlet pressure demonstrates higher accuracy, with normalized error remaining within approximately 2.5e-3p.u. In contrast, the errors of $M_{out}$ are larger, generally within 0.05p.u. This discrepancy is because the MFR is more sensitive than pressure in pipelines. Additionally, pipeline 18 (represented by the blue line), the longest pipeline in the NGS, exhibits a relatively larger error. A potential reason is that the number of delay embeddings is chosen based on the error across all pipelines, which may be insufficient for this particular pipeline. Nevertheless, the proposed globally linearized gas network model is highly accurate and suitable for system dispatch applications.

### B. Error Analysis of Different Models

In the following, we evaluate the gas flow error in the dispatch results, using the simulation results based on the nonlinear PDE model (i.e., the model (2)) under 15-minute resolution as the benchmark. To maintain consistency with the dispatch interval, we use the average hourly dispatch results and simulation results to calculate the error. We mainly focus on one type of error, the network-level error (NLE). For NLE, the gas source pressure and MFR at the gas loads from the dispatch results are treated as control variables, which are then input into the nonlinear PDE model for simulation. The errors of $P_{out}$ and $M_{in}$ between the simulation and dispatch results are calculated for all pipelines. The detailed analysis process is given in Fig. 6.

*1) Network-level error of globally linearized model*

Table II gives the error indicators for the globally linearized model under different resolutions, all at the same delay embedding orders. For the 15-minute model resolution used in this study, the MAPE is 0.04% for pressure and 0.66% for MFR, which are sufficiently small for engineering purposes. As the resolution increases, the pressure error remains low, while the MFR error rises notably due to its high sensitivity, i.e., small changes in pressure can cause great fluctuations in MFR. Despite these variations, the objective values remain consistent across all three cases. Higher resolution reduces the number of optimization variables, which in turn leads to shorter solver times.

Fig. 7 (a) and (b) show the hourly averaged normalized errors for the outlet pressure and the inlet MFR for all pipelines. The pressure error stays within 4e-3p.u., while the MFR error generally stays below 0.05p.u. This performance is consistent with the test data results in Fig. 5. Additionally, some pipelines exhibit similar error patterns, particularly neighboring ones like pipelines 18 and 19 in Fig. 7 (a). These errors may result from error propagation between pipelines.

*2) Network-level error of locally linearized model*

We utilize the stable backward Euler difference scheme in the locally linearized model. Table III shows the NLE indicators under different average flow velocities $\bar{v}$. When $\bar{v}$ ranges from 0m/s to 1m/s, the RMSE of pressure and MFR gradually decrease. However, when $\bar{v}$ increases from 1m/s to 2m/s, the RMSE rises sharply. This trend, consistent with Fig. 1, indicates that the selection of $\bar{v}$ directly influences the pressure and MFR accuracy and eventual stabilization. Overall, the smallest MAPE observed is 0.09% for pressure and 4.01% for MFR, which is relatively high for dispatch results. Despite this, variations in $\bar{v}$ has little impact on the objective value or solver time.

The error of the locally linearized model under $\bar{v} = 1$m/s for all pipelines is given in Fig. 7 (c) and Fig. 7 (d). The normalized error of $P_{out}$ is small and can be controlled within 5e-3p.u., and that of $M_{in}$ can be kept within 0.2p.u. Notably, pipelines 18 and 19 exhibit the largest pressure errors and relatively small MFR errors. This is primarily because, in network-level pipeline simulations, the pressure errors propagate from the source to the load, while MFR errors propagate in the opposite direction, from the load to the source. Additionally, errors tend to accumulate along the pipelines.

*3) Comparison of network-level error*

The NLE of the two models is compared below. As shown in Table II and Table III, the globally linearized model reduces

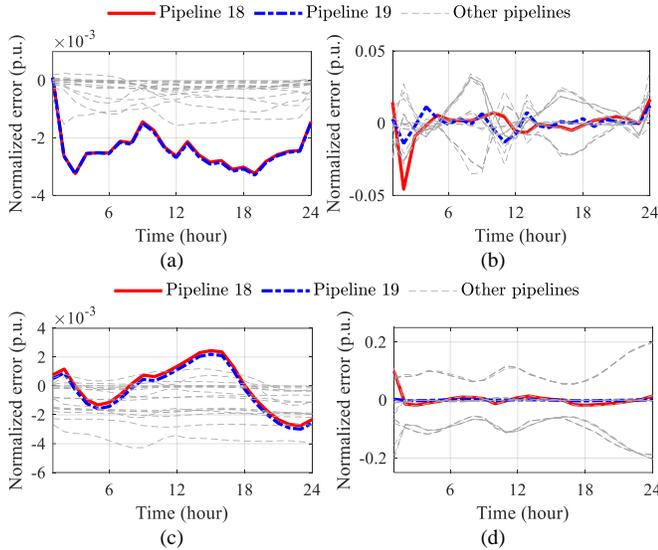

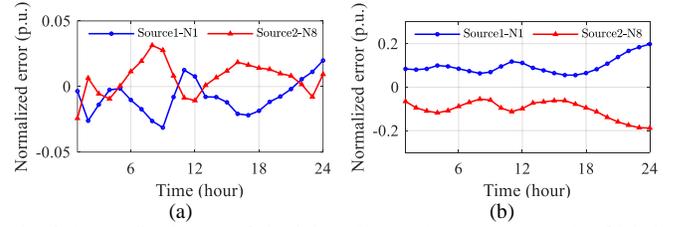

Fig. 8. Normalized error of the inlet MFR at the gas sources: (a) Globally linearized model; (b) Locally linearized model.

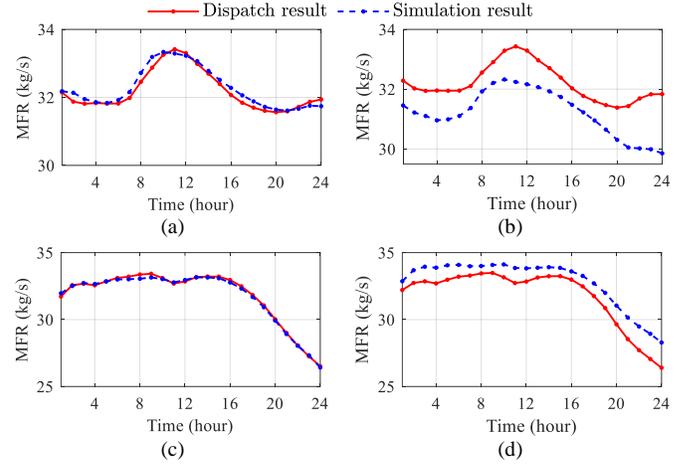

Fig. 7. The NLE of dispatch results: (a) Pressure of the globally linearized model; (b) MFR of the globally linearized model; (c) Pressure of the locally linearized model; (d) MFR of the locally linearized model.

TABLE III
Network-Level Error of Locally Linearized Model

| | $\bar{v}$ (m/s) | 0 | 1 | 2 |
|---|---|---|---|---|
| RMSE | $p_{out}$ (MPa) | 0.047226544 | 0.007643174 | 0.055132804 |
| | $M_{in}$ (kg/s) | 1.264 | 0.798 | 2.325 |
| MAPE | $p_{out}$ | 0.60% | 0.09% | 0.73% |
| | $M_{in}$ | 5.93% | 4.01% | 11.05% |

Fig. 9. The MFR of gas sources: (a) Source 1 of globally linearized model; (b) Source 1 of locally linearized model; (c) Source 2 of globally linearized model; (d) Source 2 of locally linearized model.

MAPE by 0.05% for pressure and 3.35% for MFR compared to the locally linearized model. A comparison of Fig. 7 shows that the globally linearized model exhibits more concentrated and significantly smaller errors than the locally linearized method, particularly for MFR. Additionally, most errors in Fig. 7 (c) and (d) are positive, whereas the errors in Fig. 7 (a) and (b) show more bidirectional variation. This suggests that the pipeline errors do not accumulate in a single direction but may somewhat offset each other.

Fig. 8 (a) and (b) show the inlet MFR error at gas sources. Combined with the results from Fig. 7, it becomes clear that the MFR error at the gas sources and the pressure error at the gas loads are the largest of all pipelines. The main reason is also the propagation of the error between pipelines. A detailed comparison of dispatch and simulation results for all pipelines is provided in the Appendix [28]. The prediction results of the globally linearized model are more accurate, while the error propagation in the locally linearized model is more pronounced.

The difference in objective values between the two models is minimal and can be ignored. However, the improvement in computational performance is significant. The solver time of the locally linearized model is 3.75 seconds, approximately six times longer than the 0.59 seconds required by the globally linearized model.

### C. Analysis of the Optimal Dispatch Results

The following analysis of the dispatch results further underscores the necessity of the globally linearized model. A detailed comparison of inlet MFR at the gas sources between the dispatch results and simulation results is given in Fig. 9. While the general trends of gas source dispatch results are similar for both models, the locally linearized model introduces substantially larger errors. Table IV gives the detailed error at the gas sources, showing that the maximum error of the locally linearized model at both gas sources is significantly higher than that of the globally linearized model. This large error in the locally linearized model results in a cumulative deviation of 173.71 tons during 24 hours between the planned and actual gas extraction at the sources, accounting for 3.140% of the total extraction. Such a deviation can disrupt the gas supply-demand balance. In contrast, the globally linearized model reduces the deviation by approximately 88%, accounting for only 0.384% of total extractions, thus providing more accurate and reliable dispatch decisions.

Besides, the simulation results show that using the locally linearized gas model, the MFR of three pipelines (pipeline 1, pipeline 2, and pipeline 10) in the dispatch results exceeds their operational limits, exposing critical discrepancies between the dispatch results and actual system behavior. One of the most concerning limit violations occurs in pipeline 1 connected to source 1, as shown in Fig. 9 (b). The lower limit of MFR for this pipeline is set to 30kg/s. Around the 21st hour, the MFR approaches this boundary and afterward falls below it. These violations illustrate the inadequacy of the locally linearized model in maintaining safety and operational reliability. Conversely, the globally linearized gas network model ensures that the actual system states stay within the prescribed operational limits, avoiding the risk of state violations.

In conclusion, the results show that the globally linearized model has superior performance in maintaining the operational security of the system, while the locally linearized model produces significant errors and potential security risks. Therefore, in practical applications, it is necessary to consider the nonlinear gas dynamics in the dispatch decision of the EG-IEG, for which this paper provides an effective solution.

TABLE IV
Comparison of Error at Gas Sources

| Methods | Global linear. | Local linear. |
|---|---|---|
| Max. error of source 1 (p.u.) | 0.0314 | 0.1980 |
| Max. error of source 2 (p.u.) | 0.0315 | 0.1890 |
| Accumulated error (ton) | 21.25 | 173.71 |
| Accumulated error percentage | 0.384% | 3.140% |

## VI. Conclusion

This paper proposes an optimal dispatch model for the EG-IES that considers the nonlinear gas dynamics constraints described by the PDE. Based on the Koopman operator theory, we propose a globally linearized gas network model that uses a linear system in the lifted space to describe the nonlinear gas dynamics. Besides, we introduce a Koopman operator approximation method with stability constraint to ensure the generalization capability and the accuracy of the model. Simulation results demonstrate the effectiveness and superiority of the proposed approach, which also reveal the necessity of considering the nonlinear gas dynamics in the dispatch problem of EG-IES to avoid security risk.

In this research, we choose the Koopman observables of the nonlinear gas dynamics equation based on numerical experiments, which is not a mature and scalable approach. In future research, we will explore the optimal Koopman observable selection method. Besides, we will further investigate the potentials of the globally linearized gas model in the dynamic energy flow simulation and fault propagation analysis problems of integrated energy systems.